# Insulating State of a Quasi-1-Dimensional Superconductor

J. S. Lehtinen, T. Rantala and K. Yu. Arutyunov

[1] *Department of Physics, University of Jyvaskyla, PB 35, FI-40014 Jyvaskyla, Finland*

The topic of quantum fluctuations in quasi-1D superconductors, also called quantum phase slips (QPS), has attracted a significant attention. It has been shown that the phenomenon is capable to suppress zero resistivity of ultra-narrow superconducting nanowires at low temperatures $T \ll T_c$ and quench persistent currents in tiny nanorings. It has been predicted that a superconducting nanowire in the regime of QPS is dual to a Josephson junction. In particular case of an extremely narrow superconducting nanowire embedded in high-impedance environment the duality leads to an intuitively controversial result: the superconductor enters an insulating state. Here we experimentally demonstrate that the I-V characteristic of such a wire indeed shows Coulomb blockade, which disappears with application of critical magnetic field and/or above the critical temperature proving that the effect is related to superconductivity. The system can be considered as the dynamic equivalent of a chain of conventional Josephson junctions.

Recently the subject of quasi-one-dimensional (1D) superconductivity has attracted a significant interest [1]. Remarkably, it has been demonstrated that in sufficiently narrow nanowires the basic attribute of superconductivity - zero resistivity - is not reached even at temperatures well below the critical point $T \ll T_C$ [2,3,4]. The phenomenon has been attributed to quantum fluctuations of the superconducting order parameter $\Delta = |\Delta|e^{i\varphi}$. In case of a 1D system sustaining finite supercurrent $j_S \sim |\Delta|\nabla\varphi$, the particular manifestation of the quantum fluctuations is alternatively called *quantum phase slip* (QPS), corresponding to momentary nulling of the order parameter modulus $|\Delta|$ and 'slippage' of the phase $\varphi$ by $2\pi$. In addition to finite resistivity of a 1D superconductor, the QPS effect leads to various non-trivial phenomena: for example – suppression of persistent currents in tiny superconducting nanorings [5] originating from coherent superposition of $+2\pi$ and $-2\pi$ phase slips [6].

It has been realized that in Josephson junctions (JJ) the macroscopic quantum dynamics of quasicharge $q$ and phase $\varphi$ are identical [7,8] and can be described by expressions of the same form as the Caldeira-Leggett effective action. Anticipating the equivalence of a Copper pair tunneling in a JJ and a QPS event in a superconducting nanowire, it has been pointed out [9] that Hamiltonians describing a JJ and a short superconducting nanowire in the regime of QPS, which correspondingly can be called the *quantum phase slip junction* (QPSJ), are parametrically identical. The observation reflects the fundamental quantum similarity of these two systems. For example, the substitution $E_C \leftrightarrow E_L$, $E_J \leftrightarrow E_{QPS}$, $V \leftrightarrow IR_Q$ and $\varphi \leftrightarrow \pi q/2e$, where $E_C$, $E_L$, $E_J$, $E_{QPS}$ are the energies associated with charge, inductance, Josephson and QPS couplings, $q$ is the quasicharge and $2e$ is the charge of a Cooper pair, interchanges the Hamiltonians of a voltage–biased JJ and of a current-biased QPSJ. Familiar I-V characteristic of a JJ (with critical current $I_C$) transforms to '$\pi/2$ rotated' I-V dependence (with critical voltage $V_C$) of a QPSJ. A QPSJ can be considered as the dynamic equivalent of a conventional JJ. The electrodynamics of the two systems is qualitatively indistinguishable and the extensively developed physics for JJs [10] can be 'mapped' on QPSJs. Due to the very nature of QPSJ – the absence of fixed in space and time weak link(s) – it might offer certain advantages: e.g. higher critical currents and absence of the, so-called, two level fluctuators, residing in a tunnel (Josephson) barrier. The exponential dependence of the parameter $E_{QPS}$ on the QPSJ cross section $\sigma$ enables easier, compared to conventional JJs, fabrication of QPSJs with arbitrary relation between the characteristic energies $E_C$, $E_L$ and $E_{QPS}$ [1,11]:

$$E_{QPS} = \Delta \frac{R_Q}{R_N}\left(\frac{L}{\xi}\right)^2 \exp\left(-a\frac{R_Q}{R_N}\frac{L}{\xi}\right) \quad (1)$$

where $\Delta$ is superconductor energy gap, $\xi$ - coherence length, $L$ and $R_N$ are the nanowire length and normal state resistance, $R_Q = h/(2e)^2 = 6.45$ k$\Omega$, and $a \approx 1$ is numeric parameter. For a 'dirty limit' superconductor at low temperatures $T \ll T_C$ the term under exponent in (1) reduces to $\sim -\sigma T_C^{1/2}/\rho_N$, where $\rho_N$ is normal state resistivity.

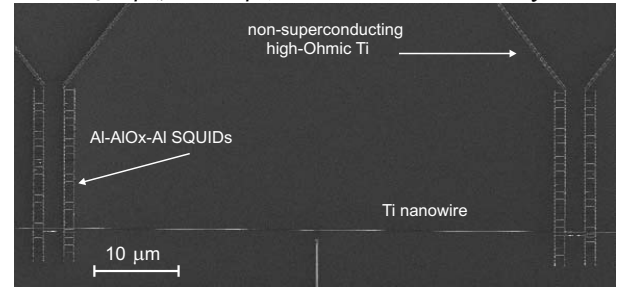

FIG. 1. SEM images of a typical nanostructure with high-impedance electrodes: 'dirty' non-superconducting titanium and 1D arrays of *Al-AlO$_x$-Al* SQUIDs.

Strictly speaking, applicability of the Eq. (1) is justified only when the QPSs are relatively rare events, e. g. the expression under the exponent should be much larger than unit [11].

In particular case of current-biased JJ (QPSJ) with large charging energy $E_C > E_J$ ($E_{QPS}$) the system can be described by equations similar to motion of an electron in periodic potential of a crystal lattice [12]. The corresponding Bloch oscillations have been observed in ultra-small JJs [13,14]. Recently similar behavior, originating from the quantum duality, has been reported in systems consisting of a superconducting central electrode ('island') isolated from the external circuit with two QPSJs – the *QPS Cooper pair transistor* [15,16]. The charge sensitivity of the device exclusively comes from coherent superposition of phase slips with two opposite directions $\pm 2\pi$ [17]. Strictly speaking, the quantum duality [7-9], has been established only for quasi-0D objects: a single JJ or a short superconducting nanowire with finite QPS rate – the QPSJ. However, very recently it has been demonstrated that the duality can be extrapolated to extended 1D objects: chains of JJs and long superconducting QPS channels [18].

The objective of this paper is to study the behavior of sufficiently long thin superconducting channels. The complementary electron transport experiments on similar nanowires [4,19], contacted by conventional low impedance probes (typically, of the same superconducting material), demonstrated broad R(T) dependencies, associated with manifestation of the QPS effect, and no signs of insulating behavior. In present study the on-chip high-impedance probes [Fig. 1] enable observation of the Coulomb effects when the quasicharge $q$, rather than the phase $\varphi$, can be considered a quasiclassical variable.

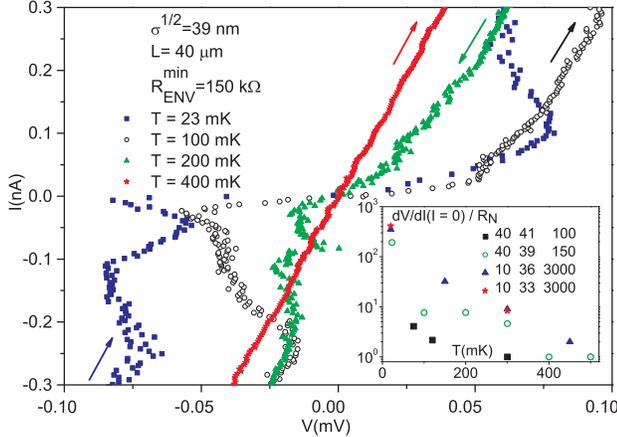

FIG. 2. I-V characteristics of the same nanowire measured at different temperatures. The quoted minimal impedance of the environment corresponds to high-bias limit $R^{min}_{ENV}=dV/dI(I\to\infty)$ determined from independent measurement of the SQUID probes. Arrows indicate direction of data recording. Inset: temperature dependencies of the zero-bias dynamic resistance of several nanowires, normalized by the normal state resistance $R_N$. Length $L$ (μm), effective diameter $\sigma^{1/2}$ (nm) and high-bias impedance of the environment $R^{min}_{ENV}$ (kΩ) are listed in the inset.

Following equation (1), titanium has been selected as the convenient material with high normal state resistivity $\rho_N$ and relatively low, but routinely achievable with a $^3$He$^4$He dilution refrigerator, critical temperature $T_C \approx 400$ mK. Conventional lift-off e-beam lithography followed by ultra-high vacuum deposition of materials were used for fabrication of the nanostructures. The homogeneity of the samples was controlled by scanning electron (SEM) and atomic force (AFM) microscopes. If necessary, further reduction of the diameter of the nanowire was obtained by low energy ion milling enabling fabrication of structures with sub-10 nm scales and with the surface roughness at the level of ± 1 nm [20]. Error in characterization of our nanowire effective diameter $\delta\sigma^{1/2}=\pm 3$ nm comes from the uncertainty in AFM determination of the interface between the metal and the substrate. The length of the samples $L$ varied from 10 μm to 50 μm. Hence, the nanowires can be considered as quasi-1D objects with the aspect ratio length/diameter ~1000. Both the $T_C$ and $\rho_N$ of titanium thin films depend on the residual pressure of foreign gases in the vacuum chamber and the deposition speed [4,16]. The observation enables fabrication of hybrid structures, where, for example, the 'body' of the sample (the QPS nanowire) is made of relatively clean titanium, while the contact probes – from 'dirty' high-Ohmic titanium, showing no traces of superconductivity down to the lowest achievable temperatures. Note that in all our structures the resistivities of both the nanowire $\rho^{QPS}_N \leq 300$ Ω/□ and the electrodes $\rho^{PROBE}_N \leq 1$ kΩ/□ are on the metallic side of metal-to-insulator transition. Coulomb effects in titanium have been observed so far in deliberately oxidized nanowires with noticeably higher resistivity [21]. An alternative approach to fabrication of high impedance on-chip electrodes [22] utilizes the concept of dynamic resistance of a JJ (or a SQUID) at zero bias $R^{DYN}_{JJ}(0)=(dI/dV)^{-1}(V\to 0)$ and/or Josephson inductance $L_J=(d^2E_J/d\Phi^2)^{-1}$. Given the conventional current-phase relation of a JJ $I_C \sim \sin(\Phi/\Phi_0)$, where $\Phi$ is the magnetic flux, and $\Phi_0$ is the superconducting flux quantum, at the degeneracy point $\Phi/\Phi_0 = \pi/2$ both the dynamic resistance and the Josephson inductance diverge. Utilization of high impedance dissipationless elements as SQUIDs [Fig. 1] has an advantage compared to dissipative high-Ohmic metallic contacts [15,16] as it eliminates the Joule heating of the latter and, correspondingly, reduces the associated Johnson noise [23]. The undesired consequence of utilization of JJs as high-impedance probes is the essentially non-linear dependence of $R^{DYN}_{JJ}$ on bias current [Fig. 4, inset]. Hereafter, the quoted environment impedance corresponds to its minimal value at high bias $R^{min}_{ENV}=dV/dI(I\to\infty)$. At small currents $I\to 0$ JJs provide orders of magnitude higher dynamic resistance [Fig. 4, inset] making them particularly efficient within the vicinity of the nanowire Coulomb blockade. All experiments were made using $^3$He$^4$He dilution refrigerator located in electromagnetically shielded room with solely battery powered analogue pre-amplifiers inside. The multi stage RLC filtering system enables reduction of the electron temperature down to ~30 mK at the base temperature of the refrigerator ~17 mK [24]. All input/output lines between the analogue front-end electronics and the rest of the measuring set-up (PC, DMMs, lock-in) were carefully filtered passing through the walls of the shielded room.

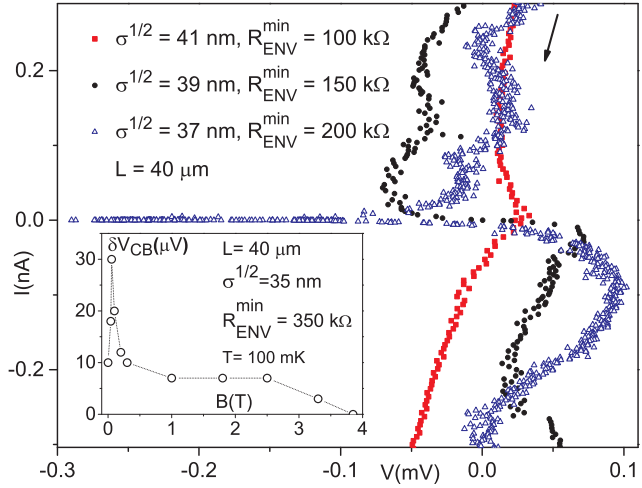

FIG. 3. Low temperature $T<<T_C$ Coulomb blockade of L=40 μm long titanium nanowire with progressively reduced cross section σ and slightly different impedance of the environment $R^{min}_{ENV}$. Arrow indicates the direction of data recording. Inset: magnetic field dependence of the Coulomb blockade gap of a similar structure. Line is guide for eyes.

Titanium nanowires with effective diameters $\sigma^{1/2} \leq 45$ nm and low-impedance electrodes have demonstrated very broad R(T) dependencies, with the thinnest samples showing marginal drop of resistance down to the lowest achievable temperatures [4,19]. The effect has been associated with manifestation of QPS effect quenching dissipationless superconducting state even at temperatures T→0. The I-V characteristics of the thickest nanowires $\sigma^{1/2} > 45$ nm demonstrated the expected (conventional) zero resistance state with well-defined critical current $I_C$. With reduction of the nanowire cross section σ the traces of the 'residual' critical current were observed in samples 30 nm $\leq \sigma^{1/2} \leq 45$ nm, while the truly zero resistance state has not been detected even at the smallest bias [4,19]. In sub-30 nm nanowires within the whole range of bias currents the dynamic resistance dV/dI was equal to the normal state resistance $R_N \equiv R(T>T_C)$ [4,19].

As those samples [4,19] were fabricated using identical technological parameters as the ones with high-impedance probes from the present study [Fig. 1], one may reasonably conclude that in the latter ones the QPS process should be present. I-V characteristics of the structures with high-impedance probes are drastically different. In sufficiently narrow nanowires (e.g. $\sigma^{1/2} \leq 40$ nm) the insulating state – Coulomb blockade – is observed [Figs. 2-4]. The effect decreases with increase of temperature and completely disappears above the critical temperature [Fig. 2] or with application of sufficiently strong magnetic field [Fig. 3, inset]. The observations support the hypothesis that the phenomenon is somehow linked to superconductivity. At a given (low) temperature $T<<T_C$ the Coulomb gap $\delta V_{CB}$ indeed strongly depends on the nanowire cross section σ [Fig. 3] following the expectation $e\delta V_{CB} \sim E_{QPS}$. In structures with close values of effective diameter $\sigma^{1/2}$ and, hence, the rate of QPSs $E_{QPS}$ the high-impedance environment favors observation of the well-defined Coulomb blockade [Fig. 4]. A certain asymmetry of the back-bending I-Vs ('Bloch nose') [Figs. 2-3] is expected [10]. Note that entering (leaving) insulating state the system switches from current (voltage) to voltage (current) biased regime. At $T<<T_C$ and I→0 the dynamic resistance of the system is extremely high dV/dI(0)→∞ being determined by the exponentially small conductivity of the parallel quasiparticle channel. The corresponding response time of the circuit is rather long (tens of minutes), and it would require excessively long experiment to obtain I-Vs with 'true' asymmetry predicted by theory [10].

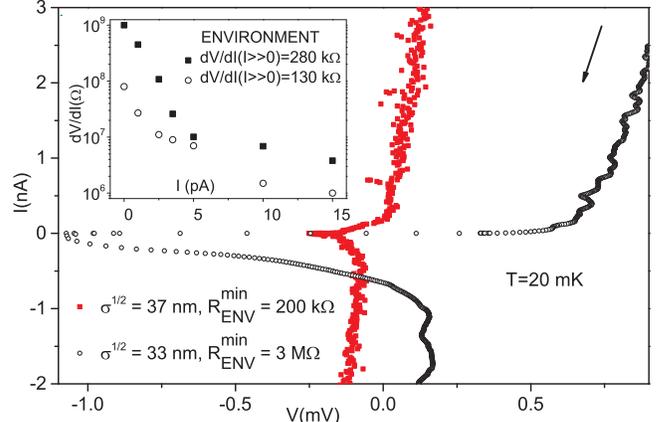

FIG. 4. Coulomb blockade of two titanium nanowires with close values of cross section σ and significantly different impedance of the environment $R^{min}_{ENV}$. Arrow indicates the direction of data recording. Inset: dynamic resistance dV/dI at various biases for typical SQUID array in series with high-resistive non-superconducting titanium wire.

One may ask a reasonable question: if our observations can account not to the QPS effect in relatively homogeneous nanowires, but rather to presence of accidentally formed JJs? As it follows from the very concept of dualism [7-9], the quantum dynamics of a JJ (or chain of JJs) and a QPSJ (or a long nanowire in the regime of QPS) is qualitatively indistinguishable. Hence, extra information (e.g. quantitative evaluation) is necessary to differentiate these two cases. All available at our disposal microscopic analyses (SEM, TEM and AFM) revealed no obvious constrictions, junctions or other sorts of weak links. In particular, TOF-ERDA studies of samples, fabricated using similar conditions, indicated ~0.4 at. % of oxygen as the highest concentration of impurities [4]. Such small amount of foreign atoms is by far insufficient to create JJ(s) across the whole cross section of a $\sigma^{1/2} = 35$ nm nanowire. Let us nevertheless estimate the contribution of such a hypothetical JJ. The maximal capacitance C, contributing to the minimal charging energy $E_C=(2e)^2/2C$, can be estimated considering a plate capacitor: 1 nm thick *TiO$_x$* barrier with dielectric constant $\varepsilon \approx 80$ separating two $\sigma^{1/2} =35$ nm electrodes. Elementary estimation gives C=0.5 fF and the corresponding $E_C / k_B = 7.2$ K ($E_C /e = 620$ μeV). The most optimistic estimation of the (maximal) Josephson energy $E_J =\hbar I_c/2e$, utilizing the values for the critical current $I_c \approx 10$ nA, extrapolated from I-Vs of homogeneous nanowires of similar cross sections [4,19], gives $E_J / k_B =0.2$ K. At $k_BT< E_J<< E_C$ the observation of Coulomb effects in such a hypothetical JJ is indeed possible [13,14]. However, then there are several problems in quantitative interpretation of our results. First, the experimental Coulomb gap $\delta V_{CB}$ depends on the cross section $\sigma$ much stronger than one would expect from such a hypothetical junction [Fig. 3]; and could be much smaller than the estimated minimal $E_C/e = 620$ μeV. Second, just above the critical temperature for titanium $T_C \approx 0.4$ K $<< E_C / k_B = 7.2$ K one might expect some traces of single electron contribution [25], which have not been detected. And finally, the suppression of the Coulomb gap by strong magnetic field at $T<<T_C$ [Fig. 3, inset] supports the relation of the effect to superconductivity and the absence of hypothetical tunnel barriers in our nanowires. It is well-known that Coulomb phenomena in single-electron systems are immune to magnetic field [26]. Note the increase of the Coulomb gap at small values of magnetic field [Fig. 3, inset]. The origin of the phenomenon is not clear and might be related to the negative magnetoresistance effect, earlier observed in QPS nanowires [3,4,19,27]. Summarizing, the association of our findings with presence of unintentionally formed JJs is rather doubtful. Certainly, the studied nanowires cannot be considered ideally homogeneous. Apparently the observed insulating state originates not from 'conventional' (static) weak links, but rather from the essentially dynamic effect – the QPS. Our data are in a reasonable accordance with estimation of the Coulomb gap $\delta V_{CB} \sim E_{QPS}$. Given the error ±3 nm in determination of the nanowire diameter and the uncertainty of the exact value of the parameter *a* in Eq. (1), establishment of a better quantitative agreement between theory [1,11] and experiment is not feasible. Insulating state of short *L*~100 nm, otherwise superconducting, *MoGe* ultrathin nanowires has been accounted to essentially normal-electron Coulomb blockade with the ratio $R_N/R_Q$ acting as the only control parameter defining the superconductor-to-insulator transition [28,29]. Already in our earlier experiments [4,19] it has been shown that in titanium nanostructures the conventional superconductivity $R(T<<T_C) \to 0$ can be merely observed in samples with $R_N>>R_Q$ (and low-Ohmic environment) if they are sufficiently 'thick' making the rate of QPS negligible in accordance with Eq. (1). There are two mandatory requirements for observation of the pronounced insulating state discussed in this *Letter*: relatively high rate of QPSs and high-impedance environment, enabling quasicharge behave as quasiclassical variable. Our observations cannot be accounted to mechanism [30], applicable to short systems [28,29] with $L<<\xi\sqrt{N_T}$, where $N_T$ is the number of normal channels in the nanowire. In addition, the temperature [Fig. 2] and magnetic field [Fig. 3, inset] dependencies of the Coulomb blockade are qualitatively different from the ones, reported in short MoGe structures [28,29].

In conclusion, we have studied the I-V characteristics of thin current-biased titanium nanowires. Earlier experiments on similar systems probed with low-impedance electrodes revealed traces of superconductivity, though being severely suppressed by quantum fluctuations of the order parameter [4,19]. Current biasing of the superconducting nanowires through on-chip high-impedance electrodes enables observation of the counterintuitive effect – the insulating state of a superconductor. The magnitude of the Coulomb gap increases with decrease of the nanowire cross section, and disappears above certain temperature and/or magnetic field supporting the relation of the effect to superconductivity. Analogous phenomena have been observed in ultra-small current-biased Josephson junctions [13,14]. The similarity originates from the fundamental quantum duality between these two systems: a Josephson junction and a quantum phase slip junction [7-9]. Observation of the Coulomb phenomena in long structures $L >> \sigma^{1/2}$ supports the universality of the dualism and its applicability to extended 1D systems [18].

The authors would like to acknowledge O. Astafiev, D. Averin, A. Bezryadin, P. Hakkonen, T. Heikkilä, L. Kuzmin, Yu. Nazarov, Yu. Pashkin, J. Pekola, A. Ustinov, A. Zaikin, and A. Zorin for valuable discussions. The work was supported by the Finnish Technical Academy project DEMAPP and National Graduate School in Material Physics NGSMP, Finland.

*Konstantin.Yu.Arutyunov@jyu.fi